\documentclass[12pt,reqno]{amsart}
\usepackage{amsmath,amsthm,amsfonts,url}
\usepackage[mathscr]{eucal}
\usepackage{eufrak}
\usepackage{graphicx}

{\obeylines
\gdef\MATH{\begingroup\parindent0pt\parskip0pt plus 0pt\obeylines%
        \def^^M{\vskip4pt}%
        \obeyspaces\tt\small}%
}
\def\goodbreakpoint{\par\penalty-5000%
         \vrule height10pt depth2pt width0pt\leavevmode}
\def\endMATH{\endgroup}
\def\MATHphi{\leavevmode
        \hbox to 0pt{\hbox to 5.24995pt{\hss$\phi$\hss}\hss}}
\def\MATHGamma{\leavevmode
        \hbox to 0pt{\hbox to 5.24995pt{\hss$\Gamma$\hss}\hss}}
\def\MATHpi{\leavevmode
        \hbox to 0pt{\hbox to 5.24995pt{\hss$\pi$\hss}\hss}}
\def\MATHinfty{\leavevmode
        \hbox to 0pt{\hbox to 5.24995pt{\hss$\infty$\hss}\hss}}
\def\MATHhStrich{\leavevmode
        \hbox to 0pt{\hbox to 5.24995pt{\vrule height4.5pt depth-3.5pt width5.24995pt}\hss}}
\def\MATHluEck{\leavevmode
        \hbox to 0pt{\hbox to 5.24995pt{\hskip2.12497pt
         \vrule height4.5pt depth1pt width1pt
         \vrule height4.5pt depth-3.5pt width2.12498pt}\hss}}
\def\MATHruEck{\leavevmode
        \hbox to 0pt{\hbox to 5.24995pt{%
         \vrule height4.5pt depth-3.5pt width2.12497pt
         \vrule height4.5pt depth1pt width1pt
         \hskip2.12498pt}\hss}}
\def\MATHloEck{\leavevmode
        \hbox to 0pt{\hbox to 5.24995pt{\hskip2.12497pt
         \vrule height9pt depth-3.5pt width1pt
         \vrule height4.5pt depth-3.5pt width2.12498pt}\hss}}
\def\MATHroEck{\leavevmode
        \hbox to 0pt{\hbox to 5.24995pt{%
         \vrule height4.5pt depth-3.5pt width2.12497pt
         \vrule height9pt depth-3.5pt width1pt
         \hskip2.12498pt}\hss}}
\def\MATHvStrich{\leavevmode
        \hbox to 0pt{\hbox to 5.24995pt{\hskip2.12497pt
         \vtop to 0pt{\hsize1pt\vss%
                \vrule height17pt depth6pt width1pt\vskip8pt\vss\par}%
         \hskip2.12498pt}\hss}}
\def\MATHtStueck{\leavevmode
        \hbox to 0pt{\hbox to 5.24995pt{%
         \vrule height4.5pt depth-3.5pt width2.12497pt
         \vrule height4.5pt depth2pt width1pt
         \vrule height4.5pt depth-3.5pt width2.12498pt}\hss}}
\def\MATHbackslash{\leavevmode
        \hbox to 0pt{\hbox to 5.24995pt{\hss$\backslash$\hss}\hss}}
\def\MATHlbrace{\leavevmode
        \hbox to 0pt{\hbox to 5.24995pt{\hss$\{$\hss}\hss}}
\def\MATHrbrace{\leavevmode
        \hbox to 0pt{\hbox to 5.24995pt{\hss$\}$\hss}\hss}}
\def\MATHkleiner{\leavevmode
        \hbox to 0pt{\hbox to 5.24995pt{\hss$\langle$\hss}\hss}}
\def\MATHkleiner{<}
\def\MATHgroesser{\leavevmode
        \hbox to 0pt{\hbox to 5.24995pt{\hss$\rangle$\hss}\hss}}
\def\MATHgroesser{>}
\def\MATHhoch{\leavevmode
        \hbox to 0pt{\hbox to 5.24995pt{\hss$^\land$\hss}\hss}}
\def\MATHtief{\leavevmode
        \hbox to 0pt{\hbox to 5.24995pt{\hss\vrule height0pt depth.8pt width3pt\hss}\hss}}
\def\n{$\tt \nu$}
\def\m{$\tt \mu$}
\def\b{$\tt \beta$}
\def\k{$\tt \kappa$}
\def\e{$\tt \varepsilon$}

\def\kk{$\tt \kappa^2$}
\def\ns{$\tt n^2$}
\def\ps{$\tt p^2$}

\allowdisplaybreaks

\theoremstyle{plain}

\numberwithin{equation}{section}
\input{tcilatex}

\begin{document}
\title[Relativistic Coulomb Integrals]
{Relativistic Coulomb Integrals and\\
Zeilberger's Holonomic Systems Approach. I}
\author{Peter Paule}
\address{Research Institute for Symbolic Computation, Johannes Kepler University,
Altenberger Stra{\ss }e 69, A-4040 Linz, Austria}
\email{Peter.Paule@risc.jku.at}
\urladdr{http://www.risc.jku.at/home/ppaule/index.html}

\author{Sergei K. Suslov}
\address{School of Mathematical and Statistics Sciences \& Mathematical,
Computational and Modeling Sciences Center, Arizona State University, Tempe,
AZ 85287-1804, U.S.A.}
\email{sks@asu.edu}
\urladdr{http://hahn.la.asu.edu/\symbol{126}suslov/index.html}
\date{June 10, 2012}
\dedicatory{Dedicated to Doron Zeilberger on
the occasion of his 60th birthday.}
\thanks{$^\dagger$Research partially supported by the Austrian Science
Foundation FWF, P20162}
\subjclass{Primary 81Q05. Secondary 33C20}
\keywords{The Dirac equation, relativistic Coulomb problem, expectation
values, generalized hypergeometric functions, holonomic systems, Gosper's algorithm, Zeilberger's fast algorithm.}

\begin{abstract}
With the help of computer algebra
we study the diagonal matrix elements $\langle Or^{p}\rangle ,$ where $O$
$=\left\{ 1, \beta , i\mathbf{\alpha n}\beta \right\} $ are the standard Dirac
matrix operators and the angular brackets denote the quantum-mechanical
average for the relativistic Coulomb problem.
Using Zeilberger's extension
of Gosper's algorithm and a variant to it, three-term recurrence relations for each of
these expectation values are derived together with some transformation formulas for
the corresponding generalized hypergeometric series.
In addition,
the virial recurrence relations for these integrals are also found
and proved algorithmically.
\end{abstract}

\maketitle

\noindent
{\sl{Science is what we understand well enough to explain to a computer. Art is everything else we do.}} \linebreak

\noindent
\hspace{14cm}{\sl{Donald~E.~Knuth \cite{A=B}}}

\section{Introduction}

This work has been initiated by the following email regarding Doron Zeilberger's Z60 conference:

\begin{center}
\url{http://www.math.rutgers.edu/events/Z60/}
\end{center}

\medskip

\noindent \textsf{\noindent {\bf{Email to Peter Paule from Sergei Suslov}}  [27 Feb 2010]}

\noindent \textsf{\noindent {\bf{Subject}}: Uranium 91+ ion}

\noindent \textsf{\textquotedblleft \dots I understand that you are coming to Doron's conference
in May and write to you with an unusual suggestion\dots}

\noindent \textsf{I am attaching two of my recent papers inspired by recent success
in checking Quantum Electrodynamics in strong fields} [see Refs.~\cite{Suslov} and \cite{SuslovC} in this paper].

\noindent \textsf{It is a very complicated problem theoretically, and fantastically,
enormously complicated (at the level of science fiction!)
experimentally, which has been solved - after 20 years of hard
work by theorists from Russia (Shabaev + 20 coauthors/students)
and experimentalists from Germany.}

\noindent \textsf{Experimentally they took a uranium 92 atom, got rid of all but
one electrons, and measured the energy shifts due to the
quantization of the electromagnetic radiation field!}

\noindent \textsf{Mathematically, among other things, the precise structure of
the energy levels of the U 91+ ion requires the evaluation of
certain relativistic Coulomb integrals, done, in a final form,
in my attached papers \dots}

\noindent \textsf{Here is the problem:}

\noindent \textsf{These integrals have numerous recurrence relations found
by physicists on the basis of virial theorems. They are also
sums of 3 (linearly dependent) 3F2 series.}

\noindent \textsf{Now you can imagine what a mess it is if one tries to derive
those relations at the level of hypergeometric series (3 times 3
= 9 functions usually!).}

\noindent \textsf{It looks as a perfect job for the G-Z algorithm in a realistic
(important) classical problem of relativistic quantum mechanics.
It looks as a good birthday present to Doron, if one could have
done that. I feel we can do that together.}

\noindent \textsf{Looking forward to your answer on my crazy suggestion,
BW, Sergei\textquotedblright}

\medskip

The first named author's computer algebra response reported at
the Z60 conference is presented in this joint paper.

\section{Relativistic Coulomb Integrals}

Recent experimental and theoretical progress has renewed interest in quantum
electrodynamics of atomic hydrogenlike systems (see, for example, \cite{Beier10},
\cite{Gum05}, \cite{Gum07}, \cite{Karsh01}, \cite{Kash03}, \cite{Mohr:Plun:Soff98}%
, \cite{ShabGreen}, \cite{ShabYFN08}, \cite{Solovyevetal10} and the references therein). In the
last decade, the two-time Green's function method of deriving formal
expressions for the energy shift of a bound-state level of high-$Z$
few-electron systems was developed \cite{ShabGreen} and numerical
calculations of QED effects in heavy ions were performed with an excellent
agreement to current experimental data \cite{Gum05}, \cite{Gum07}, \cite%
{ShabYFN08}. These advances motivate a detailed study of the expectation
values of the Dirac matrix operators multiplied by the powers of the radius
between the bound-state relativistic
Coulomb wave functions. Special cases appear in calculations of the magnetic
dipole hyperfine splitting, the electric quadrupole hyperfine splitting, the
anomalous Zeeman effect, and the relativistic recoil corrections in
hydrogenlike ions (see, for example, \cite{Adkins}, \cite{ShabHyd}, \cite%
{ShabHydVir}, \cite{Suslov} and the references therein). These expectation
values can be used in calculations with hydrogenlike wave functions when a
high precision is required. For applications of the off-diagonal matrix elements,
see \cite{Puchkov10}, \cite{Puchkov:Lab9}, \cite{Puchkov:Lab10},
\cite{ShabVest}, \cite{Shab91}, and \cite{ShabHydVir}.

\medskip
Two different forms of the radial wave functions $F$ and $G$ are available (see, for example, \cite{Ni:Uv} and \cite{Sus:Trey}). %
Given a set of parameters $a, \alpha_1, \alpha_2, \beta, \beta_1, \beta_2$, and $\gamma$, depending
on physical constants $\varepsilon, \kappa, \mu$, and $\nu$, consider
\begin{eqnarray}
\left(
\begin{array}{c}
F(r)\\
G(r)
\end{array}
\right)
= a^2 \beta^{3/2} \sqrt{ \frac{n!}{\gamma\ \Gamma(n+2\nu)} } (2 a \beta r)^{\nu -1}
e^{- a \beta r}
\left(
\begin{array}{c}
\alpha_1\ \ \alpha_2\\
\beta_1\ \ \beta_2
\end{array}
\right)
\left(
\begin{array}{c}
{L}_{n-1}^{2 \nu}(2 a \beta r)\\
{L}_{n}^{2 \nu}(2 a \beta r)
\end{array}
\right)
\end{eqnarray}
where, using the notation from \cite{DLMF}, ${L}_{n}^{\lambda}(x)$ stands for the
corresponding Laguerre polynomial of order $n$.
Throughout this paper,%
\begin{eqnarray}
&&\kappa =\pm \left( j+1/2\right) ,\qquad \nu =\sqrt{\kappa ^{2}-\mu ^{2}},
\notag \\
&&\mu =\alpha Z=Ze^{2}/\hbar c,\qquad a=\sqrt{1-\varepsilon ^{2}},
\label{notations} \\
&&\varepsilon =E/mc^{2},\qquad \beta =mc/\hbar,   \notag
\end{eqnarray}%
and $\gamma=\mu (\kappa - \nu)(\varepsilon \kappa - \nu)$,
with the total angular momentum $j=1/2,3/2,5/2$, etc.\ (see \cite{Be:Sal}, \cite{Dar}, \cite{Gor},
\cite{Schiff}, \cite{Suslov}, and
\cite{Sus:Trey} regarding the relativistic Coulomb problem).
The following identities%
\begin{eqnarray}
&&\varepsilon \mu =a\left( \nu +n\right) ,\quad \varepsilon \mu +a\nu
=a\left( n+2\nu \right) ,\quad \varepsilon \mu -a\nu =an,  \label{idents} \\
&&\varepsilon ^{2}\kappa ^{2}-\nu ^{2}=a^{2}n\left( n+2\nu \right) =\mu
^{2}-a^{2}\kappa ^{2}  \notag
\end{eqnarray}%
are useful in the calculation of the matrix elements.

The relativistic Coulomb integrals of the radial functions,%
\begin{eqnarray}
A_{p} &=&\int_{0}^{\infty }r^{p+2}\left( F^{2}\left( r\right) +G^{2}\left(
r\right) \right) \ dr,  \label{meA} \\
B_{p} &=&\int_{0}^{\infty }r^{p+2}\left( F^{2}\left( r\right) -G^{2}\left(
r\right) \right) \ dr,  \label{meB} \\
C_{p} &=&\int_{0}^{\infty }r^{p+2}F\left( r\right) G\left( r\right) \ dr,
\label{meC}
\end{eqnarray}%
have been evaluated in Refs.~\cite{Suslov} and \cite{SuslovC}
for all admissible %
integer powers $p,$ in terms of linear combinations of special generalized
hypergeometric $_{3}F_{2}$ series related to the Chebyshev polynomials of a
discrete variable \cite{Ni:Su:Uv},  \cite{Ni:Uv}.

{\it{Note.}} We concentrate on the radial integrals
since, for problems involving spherical symmetry, one can reduce all
expectation values to radial integrals by use of the properties of angular
momentum.

Throughout the paper we use the following abbreviated form of the
standard notation of the generalized
hypergeometric series $_{3}F_{2}$; see, e.g.,~\cite{DLMF}:
%
\begin{eqnarray}
 ~_{3}F_{2}\left(
\begin{array}{c}
a_1,\ a_2,\ a_3\medskip \\
b_1,\ b_2 \end{array} \right)
:= ~_{3}F_{2}\left(
\begin{array}{c}
a_1,\ a_2,\ a_3\medskip \\
b_1,\ b_2 \end{array};\ 1 \right)
= \sum_{k=0}^\infty \frac{(a_1)_k (a_2)_k (a_3)_k}{(b_1)_k (b_2)_k k!},
\end{eqnarray}
where $(a)_k:=a(a+1)\dots(a+k-1)$ denotes the Pochhammer symbol.
\medskip

%
Analogs of the traditional hypergeometric representations for the integrals are as follows \cite{Suslov}:%
\begin{eqnarray}
&&2\mu \left( 2a\beta \right) ^{p}\ \frac{\Gamma \left( 2\nu +1\right) }{%
\Gamma \left( 2\nu +p+1\right) }\ A_{p}=2p\varepsilon an~_{3}F_{2}\left(
\begin{array}{c}
1-n,\ -p,\ p\medskip +1 \\
2\nu +1,\quad 2%
\end{array}%
\right)  \label{BestA} \\
&&\qquad +\left( \mu +a\kappa \right) ~_{3}F_{2}\left(
\begin{array}{c}
1-n,\ -p,\ p+1\medskip \\
2\nu +1,\quad 1%
\end{array}%
\right) +~\left( \mu -a\kappa \right) ~_{3}F_{2}\left(
\begin{array}{c}
-n,\ -p,\ p+1\medskip \\
2\nu +1,\quad 1%
\end{array}%
\right) ,  \notag
\end{eqnarray}%
\begin{eqnarray}
&&2\mu \left( 2a\beta \right) ^{p}\ \frac{\Gamma \left( 2\nu +1\right) }{%
\Gamma \left( 2\nu +p+1\right) }\ B_{p}=2pan~_{3}F_{2}\left(
\begin{array}{c}
1-n,\ -p,\ p\medskip +1 \\
2\nu +1,\quad 2%
\end{array}%
\right)  \label{BestB} \\
&&\qquad +\varepsilon \left( \mu +a\kappa \right) ~_{3}F_{2}\left(
\begin{array}{c}
1-n,\ -p,\ p+1\medskip \\
2\nu +1,\quad 1%
\end{array}%
\right) +~\varepsilon \left( \mu -a\kappa \right) ~_{3}F_{2}\left(
\begin{array}{c}
-n,\ -p,\ p+1\medskip \\
2\nu +1,\quad 1%
\end{array}%
\right) ,  \notag
\end{eqnarray}%
\begin{eqnarray}
&&4\mu \left( 2a\beta \right) ^{p}\ \frac{\Gamma \left( 2\nu +1\right) }{%
\Gamma \left( 2\nu +p+1\right) }\ C_{p}  \label{BestC} \\
&&\qquad =a\left( \mu +a\kappa \right) ~_{3}F_{2}\left(
\begin{array}{c}
1-n,\ -p,\ p+1\medskip \\
2\nu +1,\quad 1%
\end{array}%
\right) -a\left( \mu -a\kappa \right) ~_{3}F_{2}\left(
\begin{array}{c}
-n,\ -p,\ p+1\medskip \\
2\nu +1,\quad 1%
\end{array}%
\right) .  \notag
\end{eqnarray}%
The averages of $r^{p}$ for the relativistic hydrogen atom, namely the integrals $A_p,$  were evaluated in
the late 1930s by Davis \cite{Davis} as a sum of certain three $_{3}F_{2}$
functions.\footnote{%
He finishes his article by saying: \textquotedblleft In conclusion I wish to thank Professors H.~Bateman, P.~S.~Epstein,
W.~V.~Houston, and J.~R.~Openheimer for their helpful suggestions.\textquotedblright}
But it has been realized only recently that these series are, in
fact, linearly dependent and related to the Chebyshev polynomials of a
discrete variable \cite{Suslov}.
The most compact forms in terms of only two linearly independent
generalized hypergeometric series are given in Ref.~\cite{SuslovC}.
\medskip

In addition, the integrals themselves are linearly dependent:
\begin{equation}
\left( 2\kappa +\varepsilon \left( p+1\right) \right) A_{p}-\left(
2\varepsilon \kappa +p+1\right) B_{p}=4\mu C_{p}  \label{indint1};
\end{equation}%
see, for example, \cite{Adkins}, \cite{ShabVest}, \cite{Shab91}, and \cite%
{Suslov}. Thus, eliminating, say $C_{p},$ one can deal
with $A_{p}$ and $B_{p}$ only.

The integrals (\ref{meA})--(\ref{meC}) satisfy numerous recurrence relations
in $p,$ which provide an effective way of their evaluation for small $p .$
A set of useful recurrence relations between the relativistic matrix
elements was derived by Shabaev \cite{Shab91} (see also \cite{Adkins}, \cite{Ep:Ep},
\cite{ShabVest}, \cite{ShabHydVir}, \cite{Suslov}, and \cite{Vrs:Ham})
on the basis of
a hypervirial theorem:
\begin{eqnarray}
2\kappa A_{p}-\left( p+1\right) B_{p} &=&4\mu C_{p}+4\beta \varepsilon
C_{p+1},  \label{rr1} \\
2\kappa B_{p}-\left( p+1\right) A_{p} &=&4\beta C_{p+1},  \label{rr2} \\
\mu B_{p}-\left( p+1\right) C_{p} &=&\beta \left( A_{p+1}-\varepsilon
B_{p+1}\right) .  \label{rr3}
\end{eqnarray}%
From these relations one can derive (see \cite{Adkins}, \cite{Shab91},
and \cite{ShabHydVir}) the linear relation (\ref{indint1}) and the following
computationally convenient recurrence formulas (\ref{rra})--(\ref{rrba}),
stated in our notation as
\begin{eqnarray}
A_{p+1} &=&-\left( p+1\right) \frac{4\nu ^{2}\varepsilon +2\kappa \left(
p+2\right) +\varepsilon \left( p+1\right) \left( 2\kappa \varepsilon
+p+2\right) }{4\left( 1-\varepsilon ^{2}\right) \left( p+2\right) \beta \mu }%
\ A_{p}  \label{rra} \\
&&+\frac{4\mu ^{2}\left( p+2\right) +\left( p+1\right) \left( 2\kappa
\varepsilon +p+1\right) \left( 2\kappa \varepsilon +p+2\right) }{4\left(
1-\varepsilon ^{2}\right) \left( p+2\right) \beta \mu }\ B_{p},  \notag
\end{eqnarray}%
\begin{eqnarray}
B_{p+1} &=&-\left( p+1\right) \frac{4\nu ^{2}+2\kappa \varepsilon \left(
2p+3\right) +\varepsilon ^{2}\left( p+1\right) \left( p+2\right) }{4\left(
1-\varepsilon ^{2}\right) \left( p+2\right) \beta \mu }\ A_{p}  \label{rrb}
\\
&&+\frac{4\mu ^{2}\varepsilon \left( p+2\right) +\left( p+1\right) \left(
2\kappa \varepsilon +p+1\right) \left( 2\kappa +\varepsilon \left(
p+2\right) \right) }{4\left( 1-\varepsilon ^{2}\right) \left( p+2\right)
\beta \mu }\ B_{p}  \notag
\end{eqnarray}%
and%
\begin{eqnarray}
A_{p-1} &=&\beta \frac{4\mu ^{2}\varepsilon \left( p+1\right) +p\left(
2\kappa \varepsilon +p\right) \left( 2\kappa +\varepsilon \left( p+1\right)
\right) }{\mu \left( 4\nu ^{2}-p^{2}\right) p}\ A_{p}  \label{rrab} \\
&&-\beta \frac{4\mu ^{2}\left( p+1\right) +p\left( 2\kappa \varepsilon
+p\right) \left( 2\kappa \varepsilon +p+1\right) }{\mu \left( 4\nu
^{2}-p^{2}\right) p}\ B_{p},  \notag
\end{eqnarray}%
\begin{eqnarray}
B_{p-1} &=&\beta \frac{4\nu ^{2}+2\kappa \varepsilon \left( 2p+1\right)
+\varepsilon ^{2}p\left( p+1\right) }{\mu \left( 4\nu ^{2}-p^{2}\right) }\
A_{p}  \label{rrba} \\
&&-\beta \frac{4\nu ^{2}\varepsilon +2\kappa \left( p+1\right) +\varepsilon
p\left( 2\kappa \varepsilon +p+1\right) }{\mu \left( 4\nu ^{2}-p^{2}\right) }%
\ B_{p},  \notag
\end{eqnarray}%
respectively.

{\it{Note.}} (i) These recurrences are complemented by the symmetries of the
integrals $A_{p},$ $B_{p},$ and $C_{p}$ under
the reflections $p\rightarrow -p-1$
and $p\rightarrow -p-3$ found in \cite{Suslov}; see also \cite{Andrae97}.
(ii) These relations were also derived in \cite{SuslovB} by a different
method using relativistic versions of the Kramers--Pasternack three-term recurrence
relations.


\section{Computer Algebra and Software}

The general algorithmic background of the computer algebra applications in this paper
is Zeilberger's path-breaking holonomic systems paper \cite{Zeilberger90a}. The
examples given in the following sections restrict to applications: (i) of Zeilberger's
extension \cite{Zeilberger91} of Gosper's algorithm \cite{Gosper78}, also called Zeilberger's
``fast algorithm'' \cite{Zeilberger90b, A=B}, and (ii) of a variant of it which has been described
in the unpublished manuscript \cite{PP unpublished}. Both of these algorithms have been
implemented in the Fast Zeilberger package {\tt zb.m} which is written in {\sl Mathematica}
and whose functionality is illustrated below.
A very general framework of Zeilberger's creative telescoping (i), and also of its variant (ii),
is provided by Schneider's extension of Karr's summation in difference fields \cite{Karr};
see, for instance, \cite{Schneider SLC, Schneider Para} and the references therein.


The Fast Zeilberger Package can be obtained freely from the site
\begin{center}
\url{http://www.risc.jku.at/research/combinat/software/}
\end{center}
after sending a password request to the first named author. Put the package {\tt zb.m}
in some directory, e.g., {\tt /home/mydirectory}, open a {\sl Mathematica} session,
and read in the package by
\smallskip
\MATH
\goodbreakpoint%
In[1]:= SetDirectory["/home/ppaule/RISC\_Comb\_Software\_Sep05.dir/fastZeil"];
<<zb.m
\smallskip
Fast Zeilberger Package by Peter Paule and Markus Schorn (enhanced by Axel Riese)
 - \copyright RISC Linz - V 3.53 (02/22/05)
\medskip
\goodbreakpoint%
\endMATH
A Mathematica notebook containing a full account of the Mathematica sessions described
below, together with some additional material, is available at:
\begin{center}
\url{http://hahn.la.asu.edu/~suslov/curres/index.htm}
\end{center}

\section{Unmixed Three-Term Recurrence Relations}

The following relations
purely in the $A_p$ and $B_p$, respectively, have been established in
\cite{SuslovC}:%
\begin{eqnarray}
A_{p+1} &=&\frac{\mu\, P\left(p\right) }{a^{2}\beta \left( 4\mu ^{2}\left(
p+1\right) +p\left( 2\varepsilon \kappa +p\right) \left( 2\varepsilon \kappa
+p+1\right) \right) \left( p+2\right) }\ A_{p}  \label{3termAA} \\
&&-\frac{\left( 4\nu ^{2}-p^{2}\right) \left( 4\mu ^{2}\left( p+2\right)
+\left( p+1\right) \left( 2\varepsilon \kappa +p+1\right) \left(
2\varepsilon \kappa +p+2\right) \right) p}{\left( 2a\beta \right) ^{2}\left(
4\mu ^{2}\left( p+1\right) +p\left( 2\varepsilon \kappa +p\right) \left(
2\varepsilon \kappa +p+1\right) \right) \left( p+2\right) }\ A_{p-1},  \notag
\end{eqnarray}%
\begin{eqnarray}
B_{p+1} &=&\frac{\varepsilon \mu\, Q\left(p\right) }{a^{2}\beta \left( 4\nu
^{2}+2\varepsilon \kappa \left( 2p+1\right) +\varepsilon ^{2}p\left(
p+1\right) \right) \left( p+2\right) }\ B_{p}  \label{3termBB} \\
&&-\frac{\left( 4\nu ^{2}-p^{2}\right) \left( 4\nu ^{2}+2\varepsilon \kappa
\left( 2p+3\right) +\varepsilon ^{2}\left( p+1\right) \left( p+2\right)
\right) \left( p+1\right) }{\left( 2a\beta \right) ^{2}\left( 4\nu
^{2}+2\varepsilon \kappa \left( 2p+1\right) +\varepsilon ^{2}p\left(
p+1\right) \right) \left( p+2\right) }\ B_{p-1},  \notag
\end{eqnarray}%
where%
\begin{eqnarray}
P\left( p\right) &=&2\varepsilon p\left( p+2\right) \left( 2\varepsilon
\kappa +p\right) \left( 2\varepsilon \kappa +p+1\right)  \label{PolP} \\
&&+\varepsilon \left[ 4\left( \varepsilon ^{2}\kappa ^{2}-\nu ^{2}\right)
-p\left( 4\varepsilon ^{2}\kappa ^{2}+p\left( p+1\right) \right) \right]
\notag \\
&&+\left( 2p+1\right) \left[ 4\varepsilon ^{2}\kappa +2\left( p+2\right)
\left( 2\varepsilon \mu ^{2}-\kappa \right) \right] ,  \notag \\
Q\left( p\right) &=&\left( 2p+3\right) \left[ 4\nu ^{2}+2\varepsilon \kappa
\left( 2p+1\right) +p\left( p+1\right) \right]  \label{PolQ} \\
&&-a^{2}\left( 2p+1\right) \left( p+1\right) \left( p+2\right) .  \notag
\end{eqnarray}%
In comparison with other papers
(e.g., \cite{Adkins}, \cite{Andrae97}, \cite%
{ShabVest}, \cite{Shab91}, \cite{Suslov}, \cite{SuslovB}, and the references
therein), this approach provides an alternative way of the recursive
evaluation of the special values $A_{p}$ and $B_{p},$ when one deals
separately with one of these integrals only. The corresponding initial data
$ A_{0}=1$ and $B_{-1}=a^{2}\beta /\mu $ can be found in \cite{Suslov}.

{\it Note.}
The derivation in \cite{SuslovC}
resembles the reduction (uncoupling) of the first order system of differential equations
for relativistic radial Coulomb wave functions $F$ and $G$ to the second
order differential equations;
see, for example, \cite{Ni:Uv} and \cite{Sus:Trey}.

With Zeilberger's definite extension \cite{Zeilberger90b, Zeilberger91} of Gosper's algorithm
\cite{Gosper78} for indefinite hypergeometric summation, the derivation of such recurrences
is fully automatic if the input is given as a terminating hypergeometric series (and provided
that the input is of computationally feasible size). We illustrate this by a mechanical
derivation of the following simple three-term recurrence relation for the integral $C_p$,
not found in \cite{SuslovC}:
\begin{eqnarray} \label{Crec}
C_{p+1} &=&\mu \left( 2p+1\right) \frac{2\kappa +\varepsilon \left[ p\left(
p+1\right) -4\kappa ^{2}\right] }{a^{2}\beta \left( p^{2}-4\kappa
^{2}\right) \left( p+1\right) }\ C_{p} \label{3termCC} \\
&&\ +\ p\ \frac{\left( p^{2}-4\nu ^{2}\right) \left[ \left( p+1\right)
^{2}-4\kappa ^{2}\right] }{\left( 2a\beta \right) ^{2}\left( p^{2}-4\kappa
^{2}\right) \left( p+1\right) }\ C_{p-1}.  \notag
\end{eqnarray}%
As input for $C_p$ we take the hypergeometric sum representation from \eqref{BestC}. We start
our {\sl Mathematica} session by reading in the RISC ``Fast Zeilberger'' package:
\smallskip
\MATH
\goodbreakpoint%
In[1]:= <<zb.m
\smallskip
\ \ \ \ \    Fast Zeilberger Package by Peter Paule and Markus Schorn (enhanced by Axel Riese)
\ \ \ \ \    - \copyright RISC Linz - V 3.53 (02/22/05)
\smallskip
In[2]:= {${\tt (a{\_})_{{k{\_}}}}$} := Pochhammer[a,k];
\ \ \ \ \ \ \ \ \ F1[k\_]:= {${\tt \dfrac{(1-n)_k\ (-p)_k\ (p+1)_k}{(2\ \nu+1)_k\ (1)_k\ k!}}$}; F2[k\_]:= {${\tt \dfrac{(-n)_k\ (-p)_k\ (p+1)_k}{(2\ \nu+1)_k\ (1)_k\ k!}}$};
\smallskip
In[3]:= FullSimplify[\MATHlbrace %
F1[k]/F1[k], F2[k]/F1[k]\MATHrbrace %
]
\smallskip
Out[3]= %
\MATHlbrace %
1, {${\tt -\dfrac{n}{k-n}}$}\MATHrbrace %
\goodbreakpoint%
\medskip%
In[4]:= f[k\_]:=${\tt \left(4\ \mu\ (2\ a\ \beta)^p\ \dfrac{Gamma[2\ \nu+1]}{Gamma[2\
\nu+p+1]}\right)^{-1}*F1[k]\ *\left(a\ (\mu+a\ \kappa) - a\ (\mu-a\ \kappa) * \dfrac{n}{n-k}\right);}$ %
\goodbreakpoint%
In[5]:= SuslovRec=Zb[f[k] , {k, 0, Infinity}, p, 2] // Simplify %
\medskip
Out[5]= %
\MATHlbrace %
${\tt 4\ a\ \beta\ \left(-(3+2\ p)\ ( -(2+3\ p+p^2)\ \mu^2\ (n+\nu)-2\ a\ n\ \kappa\ \mu\ (n+2\ \nu)+\right.\\}$
\ \ \ \ \ \      ${\tt \left. a^2\ \kappa^2\ (n+\nu)\ \left(2+4\ n^2+3\ p+p^2+8\ n\ \nu \right)\right)\ SUM[1+p] +\\}$
\ \ \ \ \ \ \ \    ${\tt a\ (2+p)\ \beta\ \left(-(1+p)^2\ \mu^2 + a^2\ \kappa^2\ \left(4\ n^2 + (1+p)^2 + 8\ n\ \nu \right) \right)\ SUM[p+2] == \\}$
\ \ \ \ \ \ \ \ \   ${\tt  (1+p)\ \left(1 + 2\ p + p^2 - 4\ \nu^2 \right)\ \left(-(2 + p)^2\ \mu^2 + a^2\ \kappa^2\
\left(4\ n^2 + (2 + p)^2 + 8\ n\ \nu\right) \right)\ SUM[p] \\}$\MATHrbrace %
\goodbreakpoint%
\endMATH

\noindent Here $C_p=\ ${\tt{SUM[p]}}. Utilizing two of the identities (\ref{notations})--(\ref{idents})
brings {\tt Out[5]} into the form \eqref{Crec}. In order to {\it prove} the correctness of {\tt Out[5]},
just type
\MATH
\goodbreakpoint%
In[6]:= Prove[]
\endMATH
\noindent
and the program generates automatically a pretty print version of a proof in a separate window or file,
respectively.

The computerized derivations and proofs of (\ref{3termAA})--(\ref{3termBB}) are analogous; one finds the details in
the corresponding \textsl{Mathematica} notebooks on the article's website.

\section{Related Transformations of Generalized Hypergeometric Series}

Several relations between two pairs of the generalized hypergeometric series under
consideration are given in \cite{Suslov} and \cite{SuslovC}:%
\begin{eqnarray}
&&{}_{3}F_{2}\left(
\begin{array}{c}
1-n,\ -p,\ p+1\medskip \\
2\nu +1,\quad 1%
\end{array}%
\right)  \label{L1} \\
&&\quad =\frac{\left( 2\nu +n\right) \left( 2\nu +p+1\right) \left( 2\nu
+p+2\right) \left( 2n+p+1\right) }{4\nu \left( 2\nu +1\right) \left( \nu
+n\right) \left( p+1\right) }  \notag \\
&&\qquad \qquad \times {}_{3}F_{2}\left(
\begin{array}{c}
1-n,\ p+2,\ -p-1\medskip \\
2\nu +2,\quad 1%
\end{array}%
\right)  \notag \\
&&\qquad -\frac{n\left( 4\nu +2n+p+1\right) }{2\left( \nu +n\right) \left(
p+1\right) }~_{3}F_{2}\left(
\begin{array}{c}
-n,\ p+2,\ -p-1\medskip \\
2\nu ,\quad 1%
\end{array}%
\right)  \notag
\end{eqnarray}%
and%
\begin{eqnarray}
&&{}_{3}F_{2}\left(
\begin{array}{c}
-n,\ \medskip -p,\ p+1 \\
2\nu +1,\quad 1%
\end{array}%
\right)  \label{L2} \\
&&\quad =\frac{n\left( 4\nu +2n-p-1\right) \left( 2\nu +p+1\right) \left(
2\nu +p+2\right) }{4\nu \left( 2\nu +1\right) \left( \nu +n\right) \left(
p+1\right) }~  \notag \\
&&\qquad \qquad \times {}{}_{3}F_{2}\left(
\begin{array}{c}
1-n,\ p+2,\ -p-1\medskip \\
2\nu +2,\quad 1%
\end{array}%
\right)  \notag \\
&&\qquad -\frac{\left( 2\nu +n\right) \left( 2n-p-1\right) }{2\left( \nu
+n\right) \left( p+1\right) }~_{3}F_{2}\left(
\begin{array}{c}
-n,\ p+2,\ -p-1\medskip \\
2\nu ,\quad 1%
\end{array}%
\right).  \notag
\end{eqnarray}%
In addition,
\begin{eqnarray}
&&\dfrac{p\left( p+1\right) }{2\nu +n}{}\ {}_{3}F_{2}\left(
\begin{array}{c}
1-n,\ \medskip p+1,\ -p \\
2\nu +1,\quad 2%
\end{array}%
\right)  \label{L3} \\
&&\quad =\frac{\left( p-2\nu \right) \left( 2\nu +p+1\right) }{2\left( 2\nu
+1\right) \left( \nu +n\right) }~{}{}_{3}F_{2}\left(
\begin{array}{c}
1-n,\ p+1,\ -p\medskip \\
2\nu +2,\quad 1%
\end{array}%
\right)  \notag \\
&&\qquad +\frac{\nu }{\nu +n}~_{3}F_{2}\left(
\begin{array}{c}
-n,\ p+1,\ -p\medskip \\
2\nu ,\quad 1%
\end{array}%
\right) ,  \notag
\end{eqnarray}%
and%
\begin{eqnarray}
&&\dfrac{p\left( p+1\right) }{n+2\nu }\ {}_{3}F_{2}\left(
\begin{array}{c}
1-n,\ -p,\ p\medskip +1 \\
2\nu +1,\quad 2%
\end{array}%
\right)   \notag \\
&&\ ={}_{3}F_{2}\left(
\begin{array}{c}
-n,\ -p,\ p+1\medskip  \\
2\nu +1,\quad 1%
\end{array}%
\right) -{}_{3}F_{2}\left(
\begin{array}{c}
1-n,\ -p,\ p+1\medskip  \\
2\nu +1,\quad 1%
\end{array}%
\right) .   \label{Chebyshev}
\end{eqnarray}%
These relations are ``responsible'' for the transformation between two different hypergeometric forms
of the relativistic Coulomb integral
\cite{Suslov, SuslovC}. 
The second named author was able to give only the proof of the last relation from the advanced theory of
generalized hypergeometric functions.

With the {\tt zb.m} package, one can not only prove but also find such relations, in the
literature called also contiguous relations, automatically. We illustrate this by a computer
derivation of \eqref{L1}.

\smallskip
\MATH
\goodbreakpoint%
In[1]:= <<zb.m
\smallskip
\ \ \ \ \    Fast Zeilberger Package by Peter Paule and Markus Schorn (enhanced by Axel Riese)
\ \ \ \ \    - \copyright RISC Linz - V 3.53 (02/22/05)
\smallskip
In[2]:= {${\tt (a{\_})_{{k{\_}}}}$} := Pochhammer[a,k]; \ \ \ F0[k\_] := {${\tt \dfrac{(1-n)_k\ (-p)_k\ (p+1)_k}{(2\ \nu+1)_k\ (1)_k\ k!}}$};
\ \ \ \ \ \ \ \ \  F1[k\_] := {${\tt \dfrac{(1-n)_k\ (-p-1)_k\ (p+2)_k}{(2\ \nu+2)_k\ (1)_k\ k!}}$}; F2[k\_] := {${\tt \dfrac{(-n)_k\ (-p-1)_k\ (p+2)_k}{(2\ \nu)_k\ (1)_k\ k!}}$};
\goodbreakpoint%
\endMATH
\smallskip
Our goal is to compute rational function coefficients $c_0,c_1,c_2$, free of the summation variable $k$, such that
\[
\sum_{k=0}^\infty \left( c_0 F0[k] + c_1 F1[k] + c_2 F2[k] \right) =0.
\]
With a parameterized version of Gosper's algorithm, similar to Zeilberger's extension of Gosper's algorithm,
we compute such $c_i$ together with a hypergeometric expression $g$ such that for non-negative integer $N$:
\[
\sum_{k=0}^N \left( F0[k] \left( c_0 + c_1 \frac{F1[k]}{F0[k]} + c_2 \frac{F2[k]}{F0[k]} \right) \right) = g[N].
\]
This is accomplished using the option {\tt Parameterized}; finally we send $N$ to infinity:
\smallskip
\MATH
\goodbreakpoint%
In[3]:= FullSimplify[\MATHlbrace %
1, F1[k]/F0[k], F2[k]/F0[k]\MATHrbrace %
]
\smallskip
Out[3]= %
\MATHlbrace %
1, {${\tt -\dfrac{(1+k+p)\ (1+2\ \nu)}{(-1+k-p)\ (1+k+2\ \nu)}}$},\ %
{${\tt \dfrac{n\ (1+k+p)\ (k+2\ \nu)}{2\ (k-n)\ (-1+k-p)\ \nu}}$}\MATHrbrace %
\goodbreakpoint%

\medskip%
In[4]:= Gosper[F0[k], {\MATHlbrace k, 0, N\MATHrbrace} ,
\ \ \ \ \ \ \ \ Parameterized -> %
\MATHlbrace %
1, {${\tt -\dfrac{(1+k+p)\ (1+2\ \nu)}{(-1+k-p)\ (1+k+2\ \nu)}}$}, {${\tt \dfrac{n\ (1+k+p)\ (k+2\ \nu)}{2\ (k-p)\ (-1+k-p)\ \nu}}$}\MATHrbrace ]
\smallskip
If `N' is a natural number, then:
Out[4]= %
\MATHlbrace %
${\tt \dsum\limits_{k=0}^{N} 4\ (1+p)\ \nu \ (n+\nu)\ (1+2\ \nu)\ F_{0} [k] - (1+2\ n+p)\ (n+2\ \nu)\ (1+p+2\ \nu) \\}$
\ \ \ \ \ \ \ \     ${\tt \ (2+p+2\ \nu) \ F_1[k] + \ 2\ n\ \nu \ (1+2\ \nu)\ (1+2\ n+p+4\ \nu)\ F_2[k]\ ==\\}$
\ \ \ \ \ \ \ \ \ \ \ \    ${\tt -\ ((1+N+p)\ (1+2\ \nu) \ (2\ n+4\ n^2 + n\ N + 2\ n^2\ N +3\ n\ p + 2\ n^2\ p + n\ N\ p + \\}$
\ \ \ \ \ \ \ \ \ \ \ \    ${\tt  \  n\ p^2 + 4\ \nu + 8\ n\ \nu + 8\ n^2\ \nu + 2\ N\ \nu + 6\ p\ \nu + 8\ n\ p\ \nu + 2\ N\ p\ \nu +}$
\ \ \ \ \ \ \ \ \ \ \ \    ${\tt  \ 2\ p^2\ \nu + 8\ \nu^2 + 8\ n\ \nu^2 + 8\ p\ \nu^2) \\}$
\ \ \ \ \ \ \ \ \ \ \ \    \ Pochhammer[1 - n,\ N] Pochhammer[-p,\ N] Pochhammer[1 + p,\ N])/
\ \ \ \ \ \ \ \ \ \ \ \    ${\tt \ ((1+N+2\ \nu)\ N!\ Pochhammer[1,\ N]\ Pochhammer[1 + 2\ \nu,\ N] )}$ \MATHrbrace %
\smallskip
\goodbreakpoint%
\endMATH
For $N \to \infty$ this gives the desired relation because the right hand side is $0$ when $N > p$.

The computerized proofs of (\ref{L2})--(\ref{Chebyshev}) are similar and the corresponding \textsl{Mathematica} notebooks are
available on the article's website.
%

\section{Virial Recurrence Relations}

A general procedure of verification of the linear relations between the relativistic integrals can be formulated
as follows. Start from the hypergeometric series representations for the integrals involved into
the identity/relation in question, and find all linear dependencies between the corresponding hypergeometric series
using the package {\tt zb.m}.
Substitute the integrals into the desired identity, eliminate the linear dependent sums/vectors
from this equation, and then simplify the coefficients in front of the rest of the series to zero
with the help of the standard identities among the quantum numbers of the relativistic Coulomb problem.

One can easily see that the linear relation (\ref{indint1}) is equivalent to (\ref{Chebyshev}),
and that (\ref{rr1}) follows from (\ref{indint1}) and (\ref{rr2}).
%
To illustrate our strategy, we derive (\ref{rr2}) directly from the hypergeometric representations
for the relativistic Coulomb integrals (\ref{BestA})--(\ref{BestC}). To this end, we input the
hypergeometric summands involved in the relations (\ref{BestA})--(\ref{BestC}) and (\ref{rr2}):
\smallskip
\MATH
\goodbreakpoint%
In[1]:= <<zb.m
\smallskip
\ \ \ \ \    Fast Zeilberger Package by Peter Paule and Markus Schorn (enhanced by Axel Riese)
\ \ \ \ \    - \copyright RISC Linz - V 3.53 (02/22/05)
\vfill
\eject
\newpage
\noindent
In[2]:= {${\tt (a{\_})_{{k{\_}}}}$} := Pochhammer[a,k];\ \ \  F0[k\_] := {${\tt \dfrac{(1-n)_k\ (-p)_k\ (p+1)_k}{(2\ \nu+1)_k\ (2)_k\ k!}}$};
\ \ \ \ \ \ \ \ \  F1[k\_] := {${\tt \dfrac{(1-n)_k\ (-p)_k\ (p+1)_k}{(2\ \nu+1)_k\ (1)_k\ k!}}$}; F2[k\_] := {${\tt \dfrac{(-n)_k\ (-p)_k\ (p+1)_k}{(2\ \nu+1)_k\ (1)_k\ k!}}$};
\ \ \ \ \ \ \ \ \  F3[k\_] := {${\tt \dfrac{(1-n)_k\ (-p-1)_k\ (p+2)_k}{(2\ \nu+1)_k\ (1)_k\ k!}}$}; F4[k\_] := {${\tt \dfrac{(-n)_k\ (-p-1)_k\ (p+2)_k}{(2\ \nu+1)_k\ (1)_k\ k!}}$};
\goodbreakpoint%

\smallskip
In[3]:= FullSimplify[\MATHlbrace %
1, F1[k]/F0[k], F2[k]/F0[k], F3[k]/F0[k], F4[k]/F0[k]\MATHrbrace %
]
\smallskip
Out[3]= %
\MATHlbrace %
1, {${\tt 1+k}$}, {${\tt -\dfrac{(1+k)\ n}{k-n}},\ $}%
{${\tt 1+k+ \dfrac{2\ k\ (1+k)}{1-k+p}},\ $}%
{${\tt \dfrac{(1+k)\ n\ (1+k+p)}{(k-n)\ (-1+k-p)}} $}%
\MATHrbrace %
\goodbreakpoint%

\medskip%
In[4]:= Gosper[F0[k], {\MATHlbrace k, 0, N\MATHrbrace} , Parameterized -> %
\MATHlbrace %
1, {${\tt 1+k}$}, {${\tt -\dfrac{(1+k)\ n}{k-n}},\ $} %
{${\tt 1+k+ \dfrac{2\ k\ (1+k)}{1-k+p}},\ $}
\ \ \ \ \ \ \ \    {${\tt \dfrac{(1+k)\ n\ (1+k+p)}{(k-n)\ (-1+k-p)}} $}\MATHrbrace %
 ]
\medskip
\goodbreakpoint%
If `N' is a natural number, then:
\smallskip
Out[4]= %
\MATHlbrace %
${\tt \dsum\limits_{k=0}^{N} - n\ F_{1} [k] + (1+n+p)\ F_{2} [k]  - n\  F_{3} [k] + (-1+n-p)\ F_{4} [k] == 0, \\}$
\ \ \ \ \ \ \ \  ${\tt \dsum\limits_{k=0}^{N} 2\ n\ p\ F_{0} [k] + (1+2\ n+p+2\ \nu)\ F_2[k] + (-1-p-2\ \nu)\ F_{4} [k] == \\}$
\ \ \ \ \ \ \ \  ${\tt \dfrac{2\ n\ (1+N+p)\ Pochhammer[1-n,\ N]\ Pochhammer[-p,\ N]\ Pochhammer[1+p ,\ N] }{N!\ Pochhammer[2 ,\ N]\ Pochhammer[1+2\ \nu ,\ N]}}, $
\ \ \ \ \ \ \ \  ${\tt \dsum\limits_{k=0}^{N} -2\ n\ (n+2\ \nu)\ F_{1} [k] +(1 + 2\ n + 2\ n^2 + 2\ p + 2\ n\ p + p^2 + 2\ \nu + \\}$
\ \ \ \ \ \ \ \ \ \ \ \    ${\tt 4\ n\ \nu + 2\ p\ \nu)\ F_2[k] - (1+p)\ (1+p+2\ \nu)\ F_{4} [k] ==  \\}$
\ \ \ \ \ \ \ \  ${\tt \dfrac{2\ n\ (1+N)\ (1+N+p)\ Pochhammer[1-n,\ N]\ Pochhammer[-p,\ N]\ Pochhammer[1+p ,\ N] }{N!\ %
Pochhammer[2 ,\ N]\ Pochhammer[1+2\ \nu ,\ N]}} $\MATHrbrace %
\goodbreakpoint%
\smallskip
\goodbreakpoint%
\endMATH
Notice that for ${N \to \infty}$ all the right hand sides vanish because they are $0$ when $N > p$.
\smallskip

Summarizing, the package has found the following three linear relations:%
\begin{equation}
n\left( X+U\right) -\left( 1+n+p\right) Y+\left( 1-n+p\right) V=0,
\label{lin1}
\end{equation}%
\begin{equation}
2 n p\; Z+\left( 1+2n+p+2\nu \right) Y-\left( 1+p+2\nu \right) V=0,  \label{lin2}
\end{equation}%
\begin{eqnarray}
&&2n\left( n+2\nu \right) X+(1+p)(1+p+2\nu ) V  \label{lin3} \\
\  &&\quad \ =\left[ (n+1)^{2}+2p+(n+p)^{2}+2(2n+p+1)\nu \right] Y
\notag
\end{eqnarray}%
for the following five linear dependent vectors%
\begin{equation*}
X:=\ _{3}F_{2}\left(
\begin{array}{c}
1-n,\ -p,\ p+1\medskip  \\
2\nu +1,\quad 1%
\end{array}%
\right) ,\qquad Y:=\ _{3}F_{2}\left(
\begin{array}{c}
-n,\ -p,\ p+1\medskip  \\
2\nu +1,\quad 1%
\end{array}%
\right) ,\qquad
\end{equation*}%
\begin{equation*}
Z:=\ _{3}F_{2}\left(
\begin{array}{c}
1-n,\ -p,\ p+1\medskip  \\
2\nu +1,\quad 2%
\end{array}%
\right) ,\qquad U:=\ _{3}F_{2}\left(
\begin{array}{c}
1-n,\ -p-1,\ p+2\medskip  \\
2\nu +1,\quad 1%
\end{array}%
\right) ,\qquad
\end{equation*}%
\begin{equation*}
V:=\ _{3}F_{2}\left(
\begin{array}{c}
-n,\ -p-1,\ p+2\medskip  \\
2\nu +1,\quad 1%
\end{array}%
\right) .\qquad
\end{equation*}
%
We choose to present everything in terms of $Y$ and $V$:
\smallskip
\MATH
\goodbreakpoint%
In[5]:= Lin1 = n*(X + U) - (1 + n + p)*Y + (1 - n + p)*V ;
\ \ \ \ \ \ \   Lin2 = (2 n p)*Z + (1 + 2 n + p + 2 \n)*Y - (1 + p + 2 \n)*V ;
\ \ \ \ \ \ \   Lin3 = -2 n (n + 2 \n)*X +
\ \ \ \ \ \ \ \ \ \ \ \  (1 + 2 n + 2 \ns + 2 p + 2 n p + \ps + 2 \n + 4 n \n + 2 p \n)*Y -
\ \ \ \ \ \ \ \ \ \ \ \  (1 + p) (1 + p + 2 \n)*V ;
\ \ \ \ \ \ \   Solve[Lin1==0 \&\& Lin2==0 \&\& Lin3==0, \{X, U, Z\}];
\ \ \ \ \ \ \   FullSimplify[\%] \\
\medskip
Out[5]= \MATHlbrace \MATHlbrace ${\tt X \to\ }$%
${\tt \dfrac{- (1+p)\ V\ (1+p+2\ \nu) + Y\ (2\ n^2 + 2\ n\ (1 + p + 2\ \nu) + (1 + p)\ (1 + p + 2\ \nu))}%
{2\ n\ (n+ 2\ \nu)}}, $
\goodbreakpoint%
\ \ \ \ \ \ \ \    ${\tt U \to\ }$%
${\tt \dfrac{ V\ (2\ n^2 - 2\ n\ (1 + p - 2\ \nu) + (1 + p)\ (1 + p - 2\ \nu)) - (1+p)\ Y\ (1+p-2\ \nu)}%
{2\ n\ (n+ 2\ \nu)}}, $
\goodbreakpoint%
\ \ \ \ \ \ \ \    ${\tt Z \to\ }$%
${\tt \dfrac{ V\ (1 + p + 2\ \nu) - Y\ (1+ 2\ n + p + 2\ \nu)}%
{2\ n\ p}} $ \MATHrbrace \MATHrbrace %
\goodbreakpoint%
\endMATH

\medskip
Introducing the Coulomb integrals,
\smallskip
\MATH
\goodbreakpoint%
In[6]:= Ap[X, Y, Z] := (2 \m (2 a \b)\^ (p) (Gamma[2 \n + 1])/(Gamma[2 \n + p + 1]))\^ (-1)* %
\goodbreakpoint%
\ \ \ \ \ \ \ \ \    ((\m + a \k)*X + (\m - a \k)*Y + 2 p \e a n*Z) ;
\goodbreakpoint%
\ \ \ \ \ \ \   Bp[X, Y, Z] := (2 \m (2 a \b)\^ (p) (Gamma[2 \n + 1])/(Gamma[2 \n + p + 1]))\^ (-1)* %
\goodbreakpoint%
\ \ \ \ \ \ \ \ \    (\e (\m + a \k)*X + \e (\m - a \k)*Y + 2 p a n*Z) ;
\goodbreakpoint%
\ \ \ \ \ \ \   Cplus1[U, V] := (4 \m (2 a \b)\^ (p + 1) (Gamma[2 \n + 1])/ %
\goodbreakpoint%
\ \ \ \ \ \ \ \ \    (Gamma[2 \n + p + 2]))\^ (-1)*(a (\m + a \k)*U + a (\m - a \k)*V) ;
\smallskip
\endMATH
\smallskip
\noindent
we express the desired relation in terms of $X,\dots, V$:
\smallskip
\MATH
\goodbreakpoint%
In[7]:= (2 \k)*Bp[X, Y, Z] - (p + 1)*Ap[X, Y, Z] - (4 \b)*Cplus1[U, V];
\ \ \ \      \% /. Gamma[2 + p + 2 \n] -> (1 + p + 2 \n)*Gamma[1 + p + 2 \n];
\ \ \ \      FullSimplify[\%]
\goodbreakpoint%
\smallskip
Out[7]=  ${\tt -\dfrac{1}{\mu\ Gamma[1+2\ \nu]}\ 2^{-1-p}\ (a\ \beta)^{-p}}$ \\%
\goodbreakpoint%
\ \ \ \ \ \ \      (\m ((X + Y) (1 + p - 2 \e \k) + U (1 + p + 2 \n) - V (1 + p + 2 \n)) + \\%
\goodbreakpoint%
\ \ \ \ \ \ \       a (2 n p Z (\e + p \e - 2 \k) + \k ((X - Y) (1 + p - 2 \e \k) + \\%
\goodbreakpoint%
\ \ \ \ \ \ \       U (1 + p + 2 \n) + V (1 + p + 2 \n)))) Gamma[1 + p + 2 \n]
\endMATH

\smallskip
\noindent
Next we rewrite the relevant part into a linear combination of $X,\dots, V$:

\smallskip
\MATH
\goodbreakpoint%
In[8]:= ZERO =  (\m ((X + Y) (1 + p - 2 \e \k) + U (1 + p + 2 \n) - V (1 + p + 2 \n)) + \\%
\goodbreakpoint%
\ \ \ \ \ \ \      a (2 n p Z (\e + p \e - 2 \k) + \\%
\goodbreakpoint%
\ \ \ \ \ \ \      \k ((X - Y) (1 + p - 2 \e \k) + U (1 + p + 2 \n) + V (1 + p + 2 \n)))) ;
\ \ \ \ \ \ \   Collect[ZERO,  \MATHlbrace X, Y, Z, U, V\MATHrbrace ] ;
\goodbreakpoint%
\ \ \ \ \ \ \   FullSimplify[\%]
\goodbreakpoint%
\smallskip
Out[8]= 2 a n p Z (\e + p \e - 2 \k) + Y (1 + p - 2 \e \k) (-a \k + \m) + \\
\goodbreakpoint%
\ \ \  X (1 + p - 2 \e \k) (a \k + \m) + V (a \k - \m) (1 + p + 2 \n) + U (a \k + \m) (1 + p + 2 \n)\\%
\goodbreakpoint%
\endMATH
\smallskip
\noindent Eliminating $X,$ $U$ and $Z,$
\medskip
\MATH
In[9]:= \% /. \MATHlbrace ${\tt X \to\ }$%
${\tt \dfrac{- (1+p)\ V\ (1+p+2\ \nu) + Y\ (2\ n^2 + 2\ n\ (1 + p + 2\ \nu) + (1 + p)\ (1 + p + 2\ \nu))}%
{2\ n\ (n+ 2\ \nu)}}, $
\goodbreakpoint%
\ \ \ \ \ \ \ \    ${\tt U \to\ }$%
${\tt \dfrac{ V\ (2\ n^2 - 2\ n\ (1 + p - 2\ \nu) + (1 + p)\ (1 + p - 2\ \nu)) - (1+p)\ Y\ (1+p-2\ \nu)}%
{2\ n\ (n+ 2\ \nu)}}, $
\goodbreakpoint%
\ \ \ \ \ \ \ \    ${\tt Z \to\ }$%
${\tt \dfrac{ V\ (1 + p + 2\ \nu) - Y\ (1+ 2\ n + p + 2\ \nu)}%
{2\ n\ p}} $ \MATHrbrace  ;%
\goodbreakpoint%
\ \ \ \ \ \ \   FullSimplify[\%] ;
\goodbreakpoint%
\ \ \ \ \ \ \   Collect[\%,  \MATHlbrace Y, V\MATHrbrace ] ;
\goodbreakpoint%
\ \ \ \ \ \ \   FullSimplify[\%]
\smallskip
Out[9]=  {${\tt \dfrac{(1 + p)\ V\ (1 + p + 2\ \nu)\ \left(- \mu\ (n - \varepsilon\ \kappa + \nu) + %
a\ (n^2\ \varepsilon - n\ \kappa + \varepsilon\ \kappa^2 + 2\ n\ \varepsilon\ \nu - \kappa\ \nu)\right)}{n\ (n + 2\ \nu)}}+$} \\%
\goodbreakpoint%
\ \ \ \ \ \ \   {${\tt Y\ \left((1+p-2\ \varepsilon\ \kappa)\ (-a\ \kappa + \mu) - %
\dfrac{(1+p)\ (a\ \kappa + \mu)\ (1+p-2\ \nu)\ (1+p+2\ \nu)}{2\ n\ (n + 2\ \nu)} - \right.}$} \\%
\goodbreakpoint%
\ \ \ \ \ \ \      {${\tt a\ (\varepsilon + p\ \varepsilon - 2\ \kappa) (1 + 2\ n + p + 2\ \nu) +}$} \\%
\goodbreakpoint%
\ \ \ \ \ \ \ \    {${\tt \left. \dfrac{(1+p-2\ \varepsilon\ \kappa)\ (a\ \kappa+\mu)%
\left(2\ n^2+2\ n\ (1+p+2\ \nu) + (1+p)\ (1+p+2\ \nu) \right)}{2\ n\ (n + 2\ \nu)} \right)}$} \\%
\goodbreakpoint%
\smallskip
\endMATH
\noindent
Finally, we simplify the coefficients of $V$ and $Y$:
\smallskip
\MATH
\goodbreakpoint%
In[10]:= ZeroV = - \m (n - \e \k + \n) + a (\ns \e - n \k + \e \kk + 2 n \e \n - \k \n) ;
\ \ \ \ \ \ \   ZeroY = 2 n (1 + p - 2 \e \k) (-a \k + \m) (n + 2 \n) -
\ \ \ \ \ \ \     (1 + p) (a \k + \m) (1 + p - 2 \n) (1 + p + 2 \n) -
\ \ \ \ \ \ \     2 a n (\e + p \e -2 \k) (n + 2 \n) (1 + 2 n + p + 2 \n) +
\ \ \ \ \ \ \     (1 + p - 2 \e \k) (a \k + \m) (2 \ns + 2 n (1 + p + 2 \n) + (1 + p) (1 + p + 2 \n)) ;
\ \ \ \ \ \ \   \MATHlbrace ZeroV, ZeroY\MATHrbrace  ;
\ \ \ \ \ \ \   FullSimplify[\%] ;
\ \ \ \ \ \ \   \% /. n - > (\e \m - a \n)/a ;
\ \ \ \ \ \ \   FullSimplify[\%] ;
\ \ \ \ \ \ \   \% /. \e\^ 2 - > 1 - a\^ 2 ;
\ \ \ \ \ \ \   FullSimplify[\%] ;
\ \ \ \ \ \ \   \% /. \k\^ 2 - > \n\^ 2 + \m\^ 2
%

\medskip
Out[10]= \MATHlbrace 0, 0\MATHrbrace %
\smallskip
\endMATH
\noindent
which is the name of the game.

Computerized proofs of (\ref{rr3}) and some of its extensions work the same; they are available on the article's website.

In a similar fashion, seeking for a more general linear combination of the corresponding integrals,
with the {\tt zb.m} package one can derive the following two-parameter relation:
\begin{equation}
\left[D (p+1)-C \left(2\kappa+\varepsilon(p+1)\right) \right] A_p
-\left[2 D \, \kappa-C \left(2\varepsilon\kappa+p+1 \right)\right] B_p + 4\mu C\, C_p +4\beta D\, C_{p+1}=0 ,
\end{equation}
where $C$ and $D$ are two arbitrary constants. The virial relations (\ref{indint1})--(\ref{rr2}) are its special cases.

We would like to point out the following relation:
\begin{equation}
(p+1) \left(2 \kappa C_p - \mu A_p \right)=%
\beta (p+2) \left( \varepsilon A_{p+1} - B_{p+1} \right) ,
\end{equation}
as another simple example.

{\it{Note.}} This relation is a linear combination of (\ref{rr1})--(\ref{rr3}); see \cite{Adkins}.

\section{Conclusion}

The relativistic Coulomb integrals (\ref{meA})--(\ref{meC})
were recently evaluated in a hypergeometric form \cite{Suslov}.
The corresponding system of the first order difference equations (\ref{rra})--(\ref{rrb})
has been solved in \cite{SuslovC} in terms of
linear combinations of the dual Hahn
polynomials thus providing an independent proof. Here, with the help of the Fast Zeilberger
Package {\tt zb.m} we give a direct
derivation of these results.

One of the goals of this article is to demonstrate the power
of symbolic computation for the study of relativistic Coulomb integrals.
Namely, computer algebra methods related to Zeilberger's holonomic systems approach
allow not only to verify some already known complicated
relations, but also to derive new ones without making enormously time-consuming calculations by hands
or with ad hoc usage of computer algebra procedures.

In a sequel to this article we are planning to investigate the computer-assisted derivation of recurrences,
e.g. the ``birthday recurrences'' from Section 4,
by taking as a starting point the original definition of the Coulomb integrals (\ref{meA})--(\ref{meC}).
To this end, we will use Koutschan's package {\tt HolonomicFunctions} \cite{Koutschan}.
This package, written in {\sl Mathematica},
implements further ideas related to Zeilberger's holonomic systems paradigm \cite{Zeilberger90a};
for instance, it includes implementations
of (variations of) Z's ``slow algorithm'', and algorithms by F.~Chyzak
(and B.~Salvy), and N.~Takayama. In this context we will have to exploit closure properties
of classes of special (resp.\ holonomic) sequences and functions;
an introduction to computer algebra methods for the univariate case can be found in \cite{Kauers:Paule}.

Moreover,
the {\tt zb.m} package strongly suggests that there are, in fact, four linearly independent virial recurrence relations,
see more details on the article's website,
but only three of
them (e.~g., \eqref{rr1}--\eqref{rr3})
are available in the literature.
Another next challenge is to study the off-diagonal matrix elements
that are important in applications \cite{Puchkov10}, \cite{Puchkov:Lab9}, \cite{Puchkov:Lab10}, \cite{ShabVest}, \cite{Shab91}, and \cite{ShabHydVir}. \medskip

\noindent \noindent \textbf{Acknowledgment.\/} We thank Doron Zeilberger
for valuable discussions and encouragement. We are grateful to Christian Krattenthaler
for his \LaTeX \ style file for presenting {\sl{Mathematica}} notebooks in this paper.

\end{document}